\documentclass{article}
\pdfoutput=1
\usepackage{wrapfig,lipsum,booktabs}
\PassOptionsToPackage{numbers}{natbib}
\usepackage[final]{nips_2017}
\usepackage[pdftex]{graphicx}
\usepackage{url}
\graphicspath{ {NIPSimage/} }

\usepackage[utf8]{inputenc} 
\usepackage[T1]{fontenc}    
\usepackage{hyperref}       
\usepackage{url}            
\usepackage{booktabs}       
\usepackage{amsfonts}       
\usepackage{nicefrac}       
\usepackage{microtype}      
\usepackage{amsmath}
\usepackage{subcaption}
\usepackage{ulem}

\title{Latent Space Temporal Model of Microbial Abundance to Predict Domination and Bacteremia}

\author{Ruiqi Zhong\\
\texttt{rz2383@columbia.edu} \\
$\dagger$
Department of Computer Science\\
Columbia University\\
New York, NY, United States
\And 
Tyler Joseph$^\dagger$\\
\texttt{tjoseph@cs.columbia.edu} 
\And
Joao B Xavier\\
\texttt{xavierj@mskcc.org}\\
Computational and Systems Biology Program\\
Memorial Sloan Kettering Cancer Center\\
New York, NY, United States
\And
Itsik Pe'er$^\dagger$\\
\texttt{itsik@cs.columbia.edu} 
}

\begin{document}

\maketitle

\begin{abstract}
Gut microbial composition has been linked to multiple health outcomes. 
Yet, temporal analysis of this composition had been limited to deterministic models.
In this paper, we introduce a probabilistic model for the dynamics of intestinal microbiomes that takes into account interaction among bacteria as well as external effects such as antibiotics. 
The model successfully deals with pragmatic issues such as random measurement error and varying time intervals between measurements through latent space modelling. 
We demonstrate utility of the model by using latent state features to predict the clinical events of intestinal domination and bacteremia, improving accuracy over existing methods.
We further leverage this framework to validate known links between antibiotics and clinical outcomes, while discovering new ones.
\end{abstract}

\section{Introduction}
The human intestinal environment is host to a variety of microbial organisms. This community is an essential component of human health \cite{Clemente2012}. Microbiome composition has been associated with increased risk for multiple common diseases, among them type 2 diabetes~\cite{Larsen2010}, obesity~\cite{Flint2011}, and Crohn's disease~\cite{Morgan2012}. Nonetheless, the complex interactions between host, environment, and external perturbations are just beginning to be understood \cite{Caporaso2011, David2014}.

Recent advances in metagenomic and targeted DNA sequencing, as well as reduced cost, have allowed researchers to investigate the microbiome in unprecedented detail \cite{Kuczynski2012}. However, the utility of this data in a clinical setting remains under-explored~\cite{Quigley2017}. Previous work has focused on developing deterministic or descriptive approaches to establish the role of the microbiome in providing resistance to infection, e.g. \cite{Stein2013}. Intestinal domination (an imbalance in community composition favoring one taxon) of Enterococcus, Streptococcus, or Proteobacteria were common complications, which further increased the risk of bacteremia in each patient.
Yet, these approaches ignore randomness in the underlying system as well as measurement noise, and therefore provide limited capacity to predict patient outcomes. In contrast, probabilistic approaches provide a natural way to model uncertainty, and provide a systematic way to make predictions.

Here we develop a probabilistic time-series model of microbiome community dynamics. Our goals are three-fold: i) incorporate uncertainty in the ascertainment process, ii) develop a model capable of capturing interactions within the community over time, and; iii) explicitly model external effects such as antibiotics. We test our model on a data set of 94 patients of various cancers (majority Leukemia and Lymphoma) undergoing allogenic hematopoetic stem cell transplant (allo-HSCT). Two common complications of this procedure are bacteremia, a bacterial infection of the blood, and intestinal domination, defined to be the event that relative abundance of one bacteria was greater than $30\%$. We demonstrate that by incorporating randomness into the time-series analysis of microbial community dynamics, we can more accurately predict patient outcomes such as bacteremia and intestinal domination, than using community composition alone.

\section{Methods}
\label{sec:methods}

\subsection{A Latent Probabilistic Model Based on Kalman Filters}

We observe DNA read counts from each bacterial taxon, while the true community composition --- the relative abundance of each taxon in a patient sample --- is latent. We model temporal dynamics of the community as latent factors of a linear dynamical system~\cite{Bishop2006}. The community composition at each time point depends on the previous  composition, as well as external effects such as antibiotics. The observed data are noisy observations of the latent state. Formally, our Kalman-Filter-based model ${\cal M}(A, B, Q, Q_0, R, {\bf \mu_{0})}$ is
\begin{eqnarray}
    {\bf z_0} &\sim N({\bf \mu_{0}}, Q_0) \\
    {\bf z_{t+1}} \big| {\bf z_t} &\sim N(A{\bf z_t} + B{\bf u_t}, Q) \\
    {\bf x_t} | {\bf z_t} &\sim N({\bf z_t}, R)
\end{eqnarray}
Here, time $t$ is measured in days. Tomorrow's vector representing each species' abundance, $\bf z_{t+1}$, depends on today's composition ($\bf z_t$), interactions between pairs of taxa (the matrix $A$), the antibiotic dosage ($\bf u_t$), and the effect on antibiotics on each taxon ($B$). 

Observations $\bf x_t$ pose several challenges.
First, they are not daily, nor evenly spaced, but we still model days as latent $\bf z_t$, even if  observations are available only for a subset $S$ of days.
Further challenging is data only available in relative abundances,
listing the frequency of each taxon at a time step, not the absolute count of cells.
Thus, observation vectors are constrained to the positive $L^1$ unit sphere, making time series modeling difficult. We investigated several data transformations to infer through this constraint, and decided to use the centered log-ratio transform (clr) \cite{Kurtz2015}: the vector ${\bf y} = ({\bf y}[1], ..., {\bf y}[D])$
of relative abundances of $D$ taxa, is log-transformed and mean-shifted: 
\begin{equation}
\text{clr}({\bf y}) = \left(\log \frac{{\bf y}[1]}{(\prod_{i=1}^D{\bf y}[i])^{1/D}}, ...,\log \frac{{\bf y}[D]}{(\prod_{i=1}^D{\bf y}[i])^{1/D}} \right)
\end{equation}
We map ${\bf y}$ to a modeled observation ${\bf x}\in {\mathbb R}^{D-1}$ as ${\bf x} = (\text{clr}(\textbf{y})[1],\ldots,\text{clr}(\textbf{y})[D-1])$. This mapping is invertible as 
\begin{equation}
{\bf y} = \text{normlize}( (e^{\textbf{x}[1]},\ldots,e^{\textbf{x}[D-1]}, e^{-\sum_d \textbf{x}[d]})) 
\end{equation}

We learn parameters $A, B, Q, Q_0, R, {\bf \mu_0}$ using an EM algorithm \cite{Dempster1977}.
 Given this formulation, we can still compute both marginal conditional expectations of $\bf z_t$ in the E step and closed form expressions for the optimal model parameters in the M step regardless of missing daily observations, in the transformed, unconstrained space.

\subsection{Prediction of Clinical Events}

To predict the clinical events of bacterial domination and bacteremia,
we trained our Kalman Filter model on other samples (50X cross-validation), and fit a linear logistic classifier~\cite{Pedregosa2011}. The classifier's input was latent composition vectors predicted by the model for the time of the event. 
This was compared against a baseline of predictions based on the most recent observation of a composition vector.
We repeated these predictions using composition vectors either in original frequency space, or as clr-transformed vectors.
For domination of a particular taxon, we further explored regressing only on the abundance of that taxon.

We considered several predictions of clinical events of bacterial domination and bacteremia.
 Predictions used logistic regression.
We repeated these predictions using either the clr-transformed vectors or the abundances.

For evaluating the contribution of antibiotics to clinical events, we forward simulated the model starting from each observation time point for 10 days. We reported the simulation with or without the drug under consideration and registered the reported probabilities for the clinical event.

\subsection{Data and availability} 
439 measurements of gut microbiome composition were collected across 94 subjects undergoing allo-HSCT~\cite{Taur2012} followed by a regimen of antibiotics. Measurements were spaced 1-21 days (median: 7), and spanned a period of 0-13 days (median: 6) before and 7-35 days (median: 18) after the transplantation.
Measurements were taken by extracting DNA from fecal specimens amplifying the V1-V3 region of the 16S rRNA genes, and phylogenetic classification of each sequence performed at the genus level (see~\cite{Taur2012} for more details).

\section{Results}

\begin{figure}
\includegraphics[width=\linewidth]{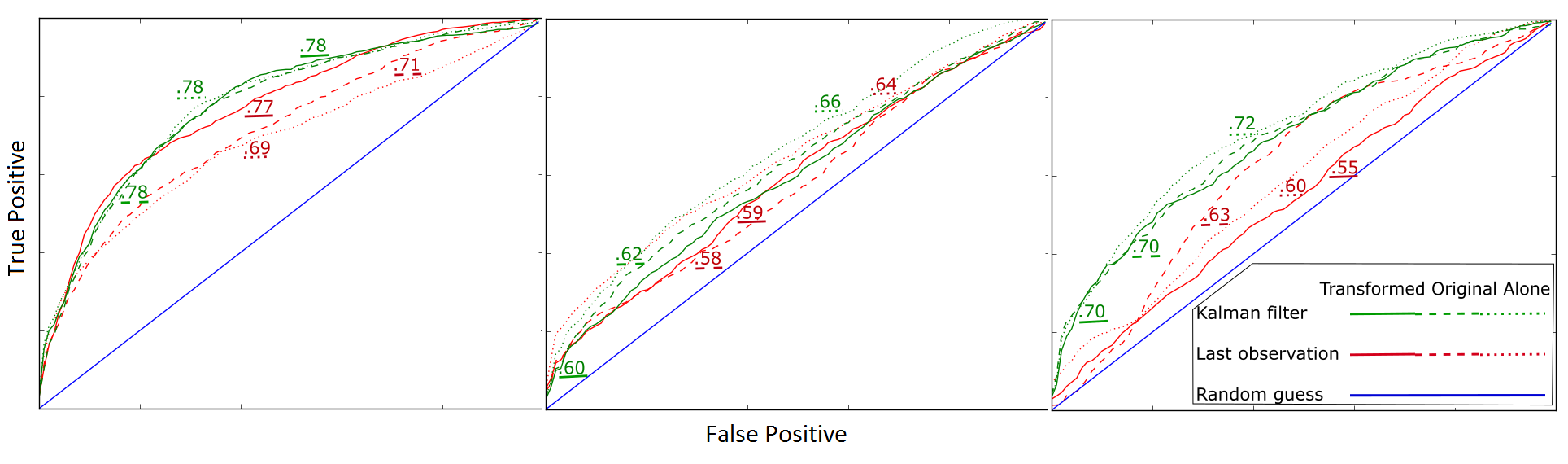}
\caption{\label{fig:pred_domination} Receiver-Operator Curves (ROC) demonstrating performance of our classifier on predicting intestinal domination by Enterococcus (left) Streptococcus (middle) and Proteobacteria (right).We compare the classifiers built by: (1) clf transformation (solid) vs. original frequency (dashed) vs. frequency of that bacteria alone (dotted) ; and (2) Kalman Filter prediction (green) vs. Most recent observation (red)}
\end{figure}

We first sought to predict which patients will develop intestinal domination of one of the above bacteria. 
Our Kalman Filter predictions of bacterial domination (Figure~\ref{fig:pred_domination}) consistently exceed baseline performance. Specifically,  Enteroccocus domination is predicted by the Kalman Filter with Area Under Curve (AUC) of 0.78, while that of most recent observation was 0.77. Similar improvements were demonstrated for Streptococcus (0.66 vs. 0.59) and  Proteobacteria domination (0.72 vs, 0.63).
Reassuringly, the best Kalman Filter prediction prediction was always the one based on the dominant taxon only.
\begin{figure}
\includegraphics[width=\linewidth]{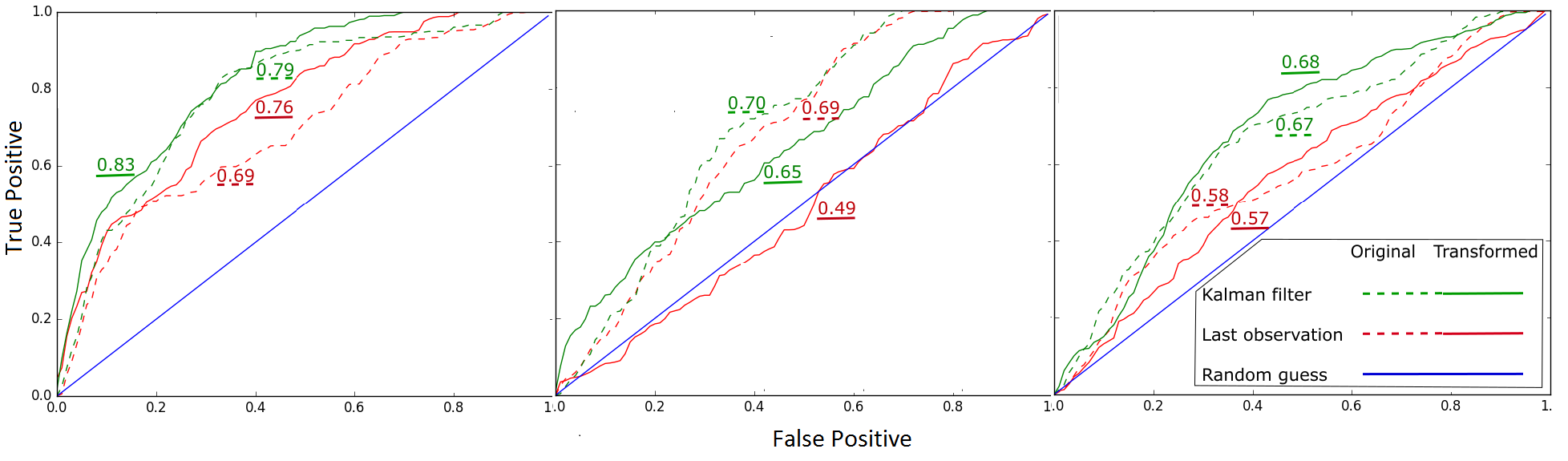}
\caption{\label{fig:pred_bacteremia}
ROCs demonstrating performance of our classifier on predicting bacteremia due to VRE (left), gram-negative bacteria (middle), or all (right). Dashed/solid and color conventions follow Figure~\ref{fig:pred_domination}.}
\end{figure}

We separately fit classifiers for the outcome variable of bacteremia, either due to Vancomycin-resistant Enterococcus (VRE), or by gram-negative bacteria.  Again, our model outperformed our baseline measures (Figure~\ref{fig:pred_bacteremia}).
AUCs for Kalman Filter predictions were 0.83 (VRE bacteremia), 0.70 (gram-negative bacteremia) and 0.68 (any bacteremia), compared to the AUCs for best MRO classifers (0.76, 0.69 and 0.58, respectively).

\begin{figure}
\begin{subfigure}{0.48\textwidth}\label{fig:dom_risk}
\includegraphics[width=\linewidth]{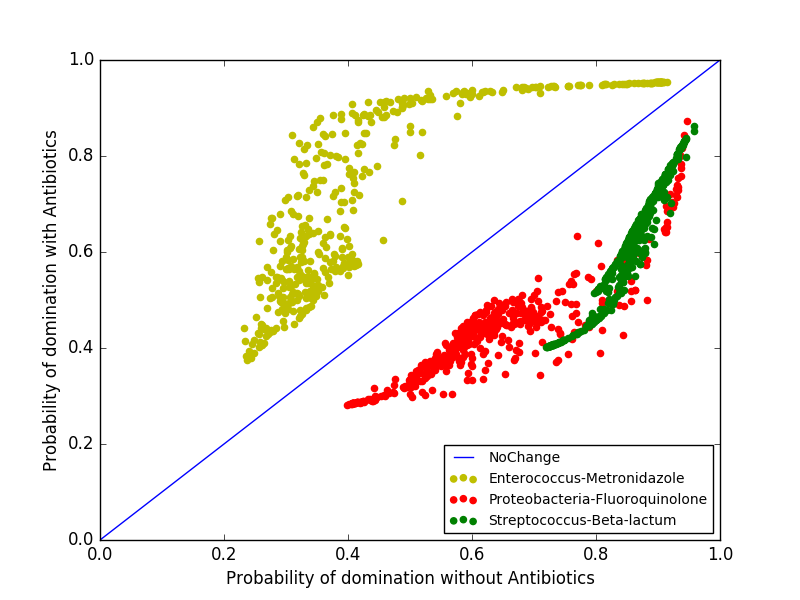}
\subcaption{}
\end{subfigure}
\begin{subfigure}{0.48\textwidth}\label{fig:bacteremia_risk} 
\includegraphics[width=\linewidth]{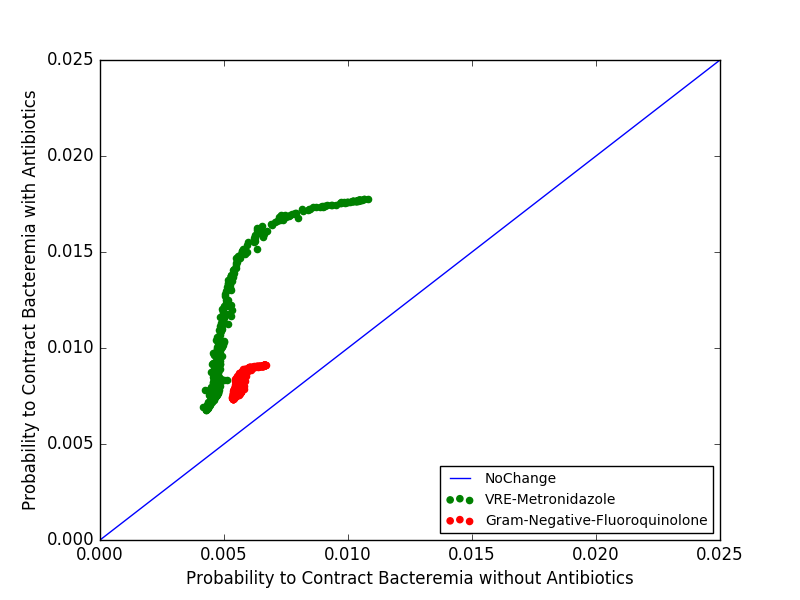}
\subcaption{}
\end{subfigure}
\caption{
\label{fig:eval_risk} 
Effect of antibiotics on intestinal domination and bacteremia. a) Metronidazole increases the risk of Enterococcus domination, Fluoroquinolone reduces the risk of Proteobacteria domination, and Beta-lactum reduces the risk of Streptococcus domination. b) Metronidazole increases the risk of VRE bacteremia and gram-negative bacteremia.}
\end{figure}

We next turned evaluating the contribution of antibiotics to bacterial domination and bacteremia (see Section~\ref{sec:methods}).
For each observation timepoint, we plot the evaluated probability of the clinical event occurrence 
with vs. without a particular antibiotic drug (Figure \ref{fig:eval_risk}). Our analysis of domination is qualitatively consistent with previous findings~\cite{Taur2012}: Metronidazole is significantly positively associated with Enterococcus domination, increasing the probability of domination by 25.7 percentage points (pp) on average and (1.6-fold); Fluoroquinolone is negatively associated with Proteobacteria domination, decreasing the probability by 19.7pp and 1.46-fold. 
We observe that these fold change evaluations are lower than those of~\cite{Taur2012}, whose respective point-estimates are 3-fold and 10-fold. We further discover a strong association between Beta-lactam and Streptococcus domination, whose domination probability it reduces by 23.7pp (1.41-fold).

An analogous analysis of bacteremia highlights Metronidazole. 
As expected from its association with Enterococcus domination,
it increases the risk of VRE bacteremia 1.9-fold. Our model further discovers  Metronidazole to be positively associated with gram-negative bacteremia, increasing its probability 1.4 fold. This is an improvement over the previous study ~\cite{Taur2012} which used survival analysis only and did not detect the effect of the beta-lactam antibiotics on the decreasing risk of Streptococcus domination.

\section{Discussion}
In this paper, we have developed a probabilistic model for microbiome community dynamics that explicitly incorporates measurement error and external effects such as antibiotics. We demonstrated the utility of this approach by applying our model to data from real patients, and showed that incorporating time-series information leads to better predictions for patient outcomes. 
Finally, we used this framework to discover links between antibiotics and clinical outcomes, validating discoveries against published results.

Kalman Filter essentially attempts to optimize estimation of the latent composition vectors, minimizing RMSE under the appropriate transformation~\cite{Kurtz2015}. It can therefore predict composition well under this transformation, but not without it (data not shown). Instead, we focused on predicting outcomes. These seem more robust to transformations, and preserve the utility of the prediction.


Nonetheless, there is still room for improvement to our model. The predicted microbiome composition, though informative, is a noisy estimate at best (data not shown). This is likely due to either data sparsity in time -- our observations are several days apart, or data size --- we only have 94 patients. Either of these could be remedied using denser data with more patients, which are likely to be available soon.

\bibliography{NIPSOct2017,NIPSTaur}
\bibliographystyle{plain}

\end{document}